\expandafter\def\csname ver@fixltx2e.sty\endcsname{}
\documentclass[conference]{IEEEtran}
\IEEEoverridecommandlockouts
\usepackage{cite}
\usepackage{amsmath,amssymb,amsfonts}
\usepackage{algorithmic}
\usepackage{graphicx}
\usepackage{dblfloatfix}
\usepackage{textcomp}
\usepackage{float}
\usepackage{multirow}
\usepackage{footnote}
\usepackage{orcidlink}
\usepackage{pgfplots}
\pgfplotsset{compat=1.18}
\usepackage{mathtools}
\usepackage[table]{xcolor}
\usepackage{pgfplotstable}
\usepackage{mathtools}
\usepackage{array}
\usepackage[nolist]{acronym}
\usepackage{pgfplots}
\pgfplotsset{compat=1.18}
\usepackage{array}
\usepackage{tabularx}
\usetikzlibrary{arrows.meta}

\usepackage{listings}
\usepackage{xcolor}
\definecolor{lightgray}{gray}{0.95}
\definecolor{darkgray}{gray}{0.3}
\definecolor{purple}{RGB}{153,0,153}

\definecolor{LightGray}{gray}{0.9}
\definecolor{LightPink}{HTML}{FFE4E1}  % Highlight color for the security penalty (Misty Rose / Light Pink)

\usepackage{booktabs}
\usepackage{amsthm}
\usepackage[textsize=tiny,colorinlistoftodos]{todonotes}
\makeatletter
\define@key{todonotes}{bh}[]{% Comment by 
	\setkeys{todonotes}{author=\textbf{Bin}, color=yellow!30}}%
\define@key{todonotes}{ak}[]{% Comment by 
	\setkeys{todonotes}{author=\textbf{Anthony}, color=blue!30}}%
\makeatother
\usepackage[normalem]{ulem}

\newif\ifreviewmode
\reviewmodetrue % Set to true to enable review mode
\reviewmodefalse

\newcolumntype{Y}{>{\centering\arraybackslash}X}

\usepackage{pgfplots}
\pgfplotsset{compat=1.18}     % shares counter with theorem
\theoremstyle{definition}

\ifreviewmode
\else
  \renewcommand{\todo}[1]{} % hide todo notes
\fi

\def\BibTeX{{\rm B\kern-.05em{\sc i\kern-.025em b}\kern-.08em
    T\kern-.1667em\lower.7ex\hbox{E}\kern-.125emX}
}

\begin{document}
    \title{Orchra: Stateful-aware Cross-slice Workload Migrations in the 6G Control Plane}

\author{
	\IEEEauthorblockN{Anthony~Kiggundu\orcidlink{0000-0003-3921-4260}\IEEEauthorrefmark{1}\IEEEauthorrefmark{2},~Bin~Han\orcidlink{0000-0003-2086-2487}\IEEEauthorrefmark{2}~and~Hans~D.~Schotten\orcidlink{0000-0001-5005-3635}\IEEEauthorrefmark{1}\IEEEauthorrefmark{2}}
	\IEEEauthorblockA{
		\IEEEauthorrefmark{1}German Research Center for Artificial Intelligence (DFKI), Germany\\
		\IEEEauthorrefmark{2}RPTU University of Kaiserslautern-Landau, Germany\\
	}
}

\maketitle

\begin{abstract}
Network slicing is a foundational capability of \ac{5G}-Advanced and emerging \ac{6G} networks, yet practical support for seamless runtime slice transitions remains limited. Standard cloud-native 5G architectures lack native support for stateful inter/intra-slice session migration, relying instead on high-overhead \ac{NAS} re-registrations, container redeployment etc., which disrupt user-plane traffic for up to $245.50\,\mathrm{ms}$.
To address this limitation, we present \textbf{Orchra}\protect\footnote{\protect\url{https://github.com/anthonyKiggundu/okra}}, an intelligent orchestrator for stateful, low-latency context transfer. By externalizing critical user equipment state—including NAS context, security keys, and \ac{PDU} session information—into a transient staging layer, Orchra preserves session continuity across slice boundaries without requiring full re-registration. 

Experimental evaluation shows that Orchra reduces this user-plane interruption by more than twice in comparison to conventional \ac{3GPP}-based approaches while incurring negligible security overhead. These results demonstrate a practical and reproducible approach for enabling seamless, state-preserving slice transitions in cloud-native \ac{5G}-Advanced networks.

\end{abstract}


\section{Introduction}
Network slicing enables multiple logical networks to coexist over a shared physical infrastructure, supporting heterogeneous services such as \ac{eMBB}, \ac{URLLC}, and \ac{mMTC}~\cite{sajjad2020interslice,afolabi2018network}. While the 3GPP architecture specifies slice selection and lifecycle management, it does not define a native mechanism for migrating an active User Equipment (UE) session between slices while preserving control- and user-plane state~\cite{3gpp.29.500,3gpp_slice_orchestra}. Consequently, open-source platforms such as \ac{OAI} and Mosaic5G implement inter-slice switching through full UE re-registration and PDU session re-establishment rather than state transfer~\cite{kaltenberger2020openairinterface,mozaic5g}. % etsi_ts122261_v16, 3gpp.23.501,

This procedure, as similarly evaluated in \cite{Sajjad} requires repeated authentication, security context establishment, and session reconstruction, introduces control-plane switching latencies that range between $110$--$250\,\mathrm{ms}$ in our OAI/Mosaic5G deployment (Section~\ref{sec:evaluation}), non-negligible user-plane interruption and packet loss. While acceptable for infrequent mobility events, this overhead limits dynamic slice adaptation driven by \ac{SLA} enforcement, edge computing, and closed-loop \ac{RAN} optimization~\cite{Mohammedali,saad2021slice}.

A further limitation is the absence of a coordinated mechanism for transferring UE context between network functions. Existing open-source implementations execute migration through loosely coupled operations across the \ac{AMF}, \ac{SMF}, and \ac{UPF} without a shared state repository or transactional coordination~\cite{mozaic5g,sakic2020decoupling}. As a result, failures during migration may require complete session re-establishment, motivating a state-persistent migration framework capable of preserving active UE context throughout slice transitions.
 
These limitations become more pronounced during inter-\ac{PLMN} cross territorial roaming, where additional signalling, authentication, and policy coordination further increase migration latency~\cite{Gallego}. Conventional Home-Routed Roaming has been reported to introduce service interruptions approaching two minutes, motivating optimized 5G Standalone roaming procedures capable of reducing interruption to the order of hundreds of milliseconds~\cite{kousaridas20215g,Marquez_etal}. Despite advances such as \ac{SEPP}, slice roaming agreements, and federated edge infrastructures, practical open-source platforms for evaluating stateful inter-slice migration remain limited~\cite{taleb2019cross,p1sec2026sepp,Nguyen_etal}.
To address these limitations, we implement a stateful slice migration framework that minimizes control-plane migration latency while preserving active UE session state.

\subsection{Novelty and Contributions}
The main contributions of this work are:
\begin{itemize}
    \item \textbf{Stateful Slice Migration:} We design a Redis-backed state externalization mechanism that preserves active UE session context across slice transitions, eliminating the need for full UE re-registration.

    \item \textbf{Network-side Stateful Migration:} We introduce an orchestration layer within the SMF that performs slice migration entirely inside the core network through direct state externalization and restoration. By avoiding UE re-registration and network-wide rediscovery procedures, \textbf{Orchra} reduces control-plane migration latency from the $112.40$--$245.50\,\mathrm{ms}$ observed in our OAI/Mosaic5G  baseline to approximately $40$--$60\,\mathrm{ms}$.

    \item \textbf{Open-source Evaluation Platform:} We have open-sourced our framework to enable reproducibility through an OAI-based testbed supporting automated performance evaluation of low-latency stateful slice migration, including user-plane continuity and cryptographic overhead measurements.
\end{itemize}

The remainder of this paper is structured as follows:
Section \ref{sec:setup} details the architecture and implementation of our orchestrator prototype, including how we integrate open tools. Section \ref{sec:results} presents our experimental setups, results, and analysis. We conclude with a discussion about some limitations, open challenges  and future directions in Section \ref{sec:conclusions}.

\section{Orchra Implementation and Setup}\label{sec:setup}

\subsection{Building the Control Plane Glue}
To support stateful UE migration from a source slice to a target slice without triggering standard 3GPP NAS re-registration loops or security renegotiations, \texttt{orchra} implements a modular inter-slice switching framework around OAI. The design is structured around three architectural components---\emph{state externalization}, \emph{control-plane interception}, and \emph{atomic user-plane re-anchoring}---which together define the migration path: UE context is temporarily externalized, the standard PDU Session Update is intercepted by a custom shim, and the target slice is brought to a ready state before the access-network switch is finalized.

\subsubsection{Active State Externalization}
The state externalization phase decouples critical UE mobility and session state variables from local network function (NF) memory, turning stateless runtime elements into persistent structural objects. This context encompasses NAS uplink/downlink counters, AMF security keys, and GPRS Tunneling Protocol User-plane (GTP-U) parameters. To mitigate the security and privacy risks associated with externalizing sensitive User Equipment (UE) and Non-Access Stratum (NAS) states, the orchestration framework embeds a security model at the deployment layer. Access control to the centralized state anchor is strictly enforced via Redis authentication configured through native Kubernetes secrets. Data in transit across the service-based architecture is secured using mutual Transport Layer Security (mTLS) transport-layer encryption to prevent plain-text exposure of session variables over the network. Finally, granular Kubernetes \texttt{NetworkPolicies} isolate the storage plane, restricting ingress traffic specifically to authorized Session Management Function (SMF) and orchestrator pods. This minimizes the cluster attack surface while preserving sub-millisecond context retrieval performance.

When an inter-slice migration is triggered—for example, following an SLA violation detected by a monitoring framework such as FlexRIC—the centralized Orchra controller invokes a proprietary state-export interface on the active source SMF (\textit{Step~1}). The source SMF serializes the current in-memory PDU session context and stores the resulting state in a shared Redis repository (\textit{Step~2}). During this state externalization phase, the source UPF continues forwarding user-plane traffic, thereby avoiding the traditional pre-migration traffic freeze while the target forwarding path is being prepared. The exported context corresponds to a single execution point within the migration workflow and is tagged with a unique \ac{UUID}-based migration identifier together with its associated timestamp. This enables deterministic tracking and correlation of each migration instance throughout the export, import, and completion phases. The source slice remains responsible for the active forwarding path until the target SMF successfully restores the exported state and the corresponding control-plane procedures complete, ensuring that incomplete migration attempts do not prematurely disrupt the existing user-plane session.

\subsubsection{Control Plane Interception via Custom Core Shims}
Once the state is anchored externally, the architecture utilizes target-side control-plane interception to reconstruct the execution environment via a \textit{Make-Before-Break} mechanism~\cite{3gpp_38_300}. This is achieved via a dedicated, low-latency communication shim embedded inside the OAI-SMF binary. 
When the Orchestrator issues a target activation trigger and retrieval directive (\textit{Step~3}) to the target SMF, the intercepting shim optimizes the initialization phase by deferring default sequential 3GPP discovery routines. Instead of awaiting the completion of standard synchronous setups, the shim concurrently intercepts the inbound PDU session update signaling and executes a low-overhead, direct memory-mapped retrieval against the external Redis state anchor. This non-blocking execution path ensures that critical session context is restored before standard signaling timeouts can be triggered, eliminating the risk of state desynchronization or data loss.

By pulling the serialized context directly from the database and injecting it into the local session management handler, the system completely bypasses standard Domain Name System (DNS) lookups, \ac{NRF} (\texttt{oai-nrf}) path selection queries, and \ac{UDR}/\ac{UDM} (\texttt{oai-udm}) authentication loops.  
Furthermore, the reconstructed context is injected directly into the target SMF without repeating the standard service discovery and subscriber retrieval procedures. As a result, the recovery path is reduced to state retrieval, validation, and \ac{PFCP} reconfiguration, minimizing the additional control-plane processing required to resume service.

\subsubsection{Atomic User-Plane Re-Anchoring}
The final phase of the migration pipeline relies on atomic user-plane re-anchoring to secure end-to-end data path continuity without packet loss. The target SMF initiates an \texttt{N4 Establish} request (Step~4) with the target UPF to prime the incoming user-plane buffers and establish downward data tunnels. Concurrently, the Orchestrator executes a coordinated \texttt{N2 Switch} directive (\textit{Step~5}) via the AMF down to the \ac{gNB}. This creates a functional routing overlap where the target slice path is completely pre-staged before the radio side alters its connection state.

As soon as the target SMF detects an inbound \texttt{resume} indication (\textit{Step~6}) from the gNB, it switches active downlink packet detection and forwarding action rules (PDRs/FARs). The target UPF immediately initiates a controlled \texttt{Flush} (\textit{Step~7}) to empty the accumulated user-plane buffer queue directly into the newly redirected RAN stream. Once this buffer is entirely drained and synchronization is finalized, the target UPF issues an asynchronous \texttt{Flush Callback} (\textit{Step~8}) back to the centralized Orchestrator to confirm zero-data-loss user-plane alignment. 

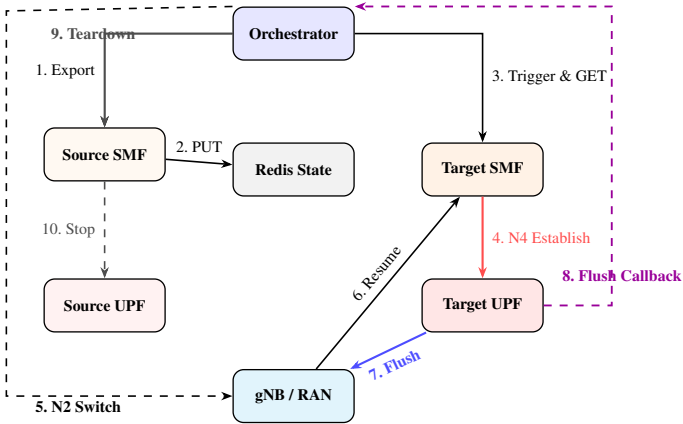
\begin{figure}[htbp]
    \centering
\begin{tikzpicture}[
   font=\fontsize{6}{7}\selectfont,
   box/.style={draw, rounded corners, align=center, minimum width=1.6cm, minimum height=0.7cm, line width=0.5pt},
   arrow/.style={-{Stealth[length=1.5mm]}, line width=0.6pt},
   dashedarrow/.style={-{Stealth[length=1.5mm]}, dashed, line width=0.6pt}
 ]

 \node[box, fill=blue!10] (orch) at (0, 4.8) {\textbf{Orchestrator}};
 \node[box, fill=gray!10] (redis) at (0, 3.0) {\textbf{Redis State}};

 \node[box, fill=orange!5] (smf_s) at (-2.5, 3.2) {\textbf{Source SMF}};
 \node[box, fill=red!5] (upf_s) at (-2.5, 1.2) {\textbf{Source UPF}};

 \node[box, fill=orange!10] (smf_t) at (2.5, 3.0) {\textbf{Target SMF}};
 \node[box, fill=red!10] (upf_t) at (2.5, 1.2) {\textbf{Target UPF}};

 \node[box, fill=cyan!10] (gnb) at (0, 0) {\textbf{gNB / RAN}};

 % --- 1. State Capture Leg ---
 \draw[arrow] (orch) -| (smf_s) node[pos=0.7, left] {1. Export};
 \draw[arrow] (smf_s) -- (redis) node[midway, above] {2. PUT};

 % --- 2. Target Path Activation Leg (Make-Before-Break) ---
 \draw[arrow] (orch) -| (smf_t) node[pos=0.7, right] {3. Trigger \& GET};
 \draw[arrow, line width=0.8pt, red!70] (smf_t) -- (upf_t) node[midway, right] {4. N4 Establish};

 % --- 3. Control Plane Completion (N2 Switch) ---
 \coordinate (n2left) at (-3.8, 5.1);
 \coordinate (n2down) at (-3.8, 0.0);
 \draw[dashedarrow]
   (orch.north west) -- (n2left)
   -- (n2down)
   -- (gnb.west)
   node[pos=0.55, left, yshift=-4] {\textbf{5. N2 Switch}};

 % --- 4. User Plane Resumption (Arrow corrected to point INBOUND to Target SMF) ---
 \draw[arrow] (gnb) -- (smf_t) node[midway, sloped, above] {6. Resume};
 \draw[arrow, line width=0.8pt, blue!70] (upf_t) -- (gnb) node[midway, below, sloped] {\textbf{7. Flush}};

 % --- 5. The Flush Callback Verification ---
 \draw[dashedarrow, color=purple, line width=0.7pt] 
   (upf_t.east) -- ++(0.9,0) |- (orch.north east)
   node[pos=0.25, left, xshift=30pt, yshift=-45pt] {\textbf{8. Flush Callback}};

 % --- 6. Clean Decommissioning ---
 \draw[arrow, color=darkgray] (orch) -| (smf_s) node[pos=0.3, left, xshift=-4pt] {\textbf{9. Teardown}};
 \draw[dashedarrow, color=darkgray] (smf_s) -- (upf_s) node[midway, left] {10. Stop};

 \end{tikzpicture}
    \caption{\footnotesize \textbf{The Orchra Migration Loop:} Source-side export is separated from target-side injection. The orchestrator uses Redis to stage UE state, then completes the path switch after the target control-plane and user-plane context are ready.}
    \label{fig:migration_compact}
\end{figure}

Upon receiving this validation callback, the Orchestrator safely executes a final \texttt{Teardown} command (\textit{Step~9}) to the source SMF, which subsequently issues a \texttt{Stop} command (\textit{Step~10}) to decommission the legacy source UPF processing loops. This deferred decommissioning strategy realizes a true Control and User Plane Separation (CUPS) primitive that entirely eliminates the transient packet-drop windows typical of standard 5G slice modifications.
Figure~\ref{fig:migration_compact} summarizes this workflow. The left side of the figure shows source-side export and shutdown, while the right side shows target-side retrieval, re-anchoring, and resumption. Together, these steps implement Orchra’s low latency migration path and preserve continuity of service for latency-sensitive MEC and URLLC-style workloads.

\section{Performance Evaluation and Results}\label{sec:results}
\subsection{Experimental Platforms}

The experimental evaluation was conducted using two Kubernetes-based 5G deployments built on the OpenAirInterface (OAI) core network and ueransim. Both deployments execute on the same bare metal \textit{Ubuntu 22.xx} installation, ensuring identical hardware resources, network topology, and software environment. Consequently, any observed performance differences are attributable solely to the migration mechanism rather than variations in the underlying infrastructure.

\subsubsection{Baseline Deployment}

The baseline platform implements a conventional network slicing architecture using the Mosaic5G slice controller. It comprises a complete OpenAirInterface (OAI) 5G core network, ueransim-based radio access components, and a MySQL backend. The software components deployed in the baseline environment are summarized in Table~\ref{tab:deployment_components}.

\begin{table*}[t]
    \centering
    \caption{\footnotesize Software Components in the Experimental Deployments}
    \label{tab:deployment_components}
    \rowcolors{3}{LightGray}{white}
    \scriptsize
    \begin{tabularx}{\textwidth}{l|ccccccc|cc|ccc}
        \toprule
        & \multicolumn{7}{c|}{\textbf{OAI 5G Core Functions}} & \multicolumn{2}{c|}{\textbf{RAN / Storage}} & \multicolumn{3}{c}{\textbf{Control \& Orchestration}} \\
        \textbf{Deployment} & \textbf{AMF} & \textbf{SMF} & \textbf{UPF} & \textbf{NRF} & \textbf{AUSF} & \textbf{UDM} & \textbf{UDR} & \textbf{MySQL} & \textbf{ueransim} & \textbf{Mosaic5G} & \textbf{Redis} & \textbf{Orchra} \\
        \midrule
        \textbf{Baseline} & \checkmark & \checkmark & \checkmark & \checkmark & \checkmark & \checkmark & \checkmark & \checkmark & \checkmark & \checkmark & -- & -- \\
        \textbf{Orchra}   & \checkmark & \checkmark & \checkmark & \checkmark & \checkmark & \checkmark & \checkmark & \checkmark & \checkmark & -- & \checkmark & \checkmark \\
        \bottomrule
    \end{tabularx}
\end{table*}

\subsubsection{Orchra Deployment}

The Orchra deployment retains the same OAI core network functions and ueransim components while replacing the Mosaic5G slice controller with the proposed Orchra orchestration service. A Redis state store is additionally introduced to externalize and restore active SMF session contexts during migration. The deployed software components are likewise summarized in Table~\ref{tab:deployment_components}.

\subsection{Experimental Workflow}
In the baseline deployment, slice switching follows the conventional OAI/Mosaic5G procedure, requiring the standard control-plane signalling necessary to establish service on the destination slice.
Conversely, Orchra performs stateful migration by exporting the active SMF session context to Redis, importing the stored context into the destination SMF, updating PFCP state at the UPF, and completing the required N2 signalling procedures without requiring complete NAS re-registration or PDU session re-establishment.

To evaluate migration under realistic traffic conditions, a continuous \ac{UDP} stream is generated on the \texttt{ueransim-ue} pod using \texttt{iperf3} operating at $50~Mbps$ with a reporting interval of $10~ms$. Traffic remains active throughout every migration, allowing user-plane interruption, packet loss, and jitter to be measured directly during slice transitions.
We conducted $N=200$ migration iterations, alternating between \textit{eMBB $\rightarrow$ URLLC} and \textit{URLLC $\rightarrow$ eMBB}. All measurements are collected automatically and exported into a unified CSV dataset for our subsequent statistical analysis (Section~\ref{sec:evaluation}).
% \begin{itemize}
%     \item 
%     \item 
% \end{itemize}
\subsection{Measurement Methodology}
The evaluation distinguishes between three latency components: \emph{control-plane reconfiguration latency}, \emph{user-plane recovery latency}, and the resulting \emph{end-to-end switching delay}. Separating these metrics allows us to quantify the time spent performing control-plane state migration independently from the residual delay before user traffic resumes on the destination UPF. Unlike conventional slice switching, Orchra eliminates repeated NAS registration and PDU session establishment, leaving only the control-plane synchronization and forwarding convergence required to resume service.
The following timestamps are recorded during each migration:
\begin{itemize}
    \item $t_{\mathrm{trigger}}$: migration request issued by the orchestrator;
    \item $t_{\mathrm{export}}$: source SMF completes session state export;
    \item $t_{\mathrm{import}}$: destination SMF completes state restoration;
    \item $t_{\mathrm{pfcp}}$: UPF acknowledges PFCP session modification;
    \item $t_{\mathrm{n2}}$: AMF completes the N2 tunnel update, marking the end of the control-plane migration;
    \item $t_{\mathrm{first\_pkt}}$: first user packet successfully received on the destination GTP-U tunnel.
\end{itemize}

The individual state migration phases are computed as
\begin{equation}
    %\small
    \begin{aligned}
    t_{\mathrm{export}}
        &= t_{\mathrm{export,end}}
       - t_{\mathrm{trigger}},\\
    t_{\mathrm{import}}
        &= t_{\mathrm{import,end}}
       - t_{\mathrm{import,start}},\\
    t_{\mathrm{pfcp}}
        &= t_{\mathrm{pfcp,end}}
       - t_{\mathrm{pfcp,start}},\\
    t_{\mathrm{n2}}
        &= t_{\mathrm{n2,end}}
       - t_{\mathrm{n2,start}}.
    \end{aligned}
\end{equation}
The aggregate migration metrics are defined as
\begin{equation}
\begin{aligned}
t_{\mathrm{CP}}
    &= t_{\mathrm{export}}
     + t_{\mathrm{import}}
     + t_{\mathrm{pfcp}}
     + t_{\mathrm{n2}}, \\
t_{\mathrm{UP}}
    &= t_{\mathrm{first\_pkt}}
     - t_{\mathrm{n2,end}}, \\
t_{\mathrm{switch}}
    &= t_{\mathrm{first\_pkt}}
     - t_{\mathrm{trigger}} \\
    &= t_{\mathrm{CP}}
     + t_{\mathrm{UP}}.
\end{aligned}
\end{equation}

Here, $t_{\mathrm{export}}$ denotes the time required to serialize and externalize the session state at the source SMF, $t_{\mathrm{import}}$ captures the time required to retrieve and restore the session state at the destination SMF, $t_{\mathrm{pfcp}}$ represents the PFCP session reconfiguration latency over the N4 interface, and $t_{\mathrm{n2}}$ measures the duration of the N2 signaling required to complete the control-plane update. The aggregate control-plane migration latency, $t_{\mathrm{CP}}$, is the sum of these four phases. The user-plane recovery latency, $t_{\mathrm{UP}}$, measures the additional time required for the first packet to be received after control-plane completion, while $t_{\mathrm{switch}}$ represents the total end-to-end service interruption experienced during slice migration.

\noindent For each metric, results are collected over $N=200$ independent migration runs and summarized using the mean, standard deviation, and 95\% confidence interval.

\subsection{Security Architecture and Cryptographic Overhead Evaluation}
The proposed framework employs a two-layer security architecture to protect externally stored session state during inter-slice migration. The first layer secures the Redis state repository using standard Kubernetes mechanisms, including namespace isolation, access control, and secret management. As these deployment-level protections introduce negligible runtime overhead, they are excluded from the performance evaluation.

Beyond the infrastructure-level protections, \textbf{Orchra} secures externally stored session state using application-layer authenticated encryption. Under the baseline \emph{plaintext} migration path, the source SMF serializes the active PDU session context into a JSON payload and stores it directly in Redis. Under the \emph{secured} migration path, the serialized state is encrypted and authenticated using the Advanced Encryption Standard operating in Galois/Counter Mode (\ac{AES-GCM}) with a 256-bit key before being externalized. Upon retrieval, the destination SMF verifies the authentication tag, decrypts the stored state, and reconstructs the active session context.

Since cryptographic serialization and deserialization are performed for every migration, we isolate these micro-operations to quantify their contribution to the state export ($t_{\mathrm{export}}$) and state import ($t_{\mathrm{import}}$) latency relative to the plaintext state migration baseline.

\subsection{Evaluation Results}\label{sec:evaluation}
Table~\ref{table:evaluation_metrics} compares three network slice switching strategies: full pod recreation, process-level restart without pod recreation, and Orchra-assisted state migration. Results are reported as the mean, standard deviation, and $95\%$ confidence interval over $N=200$ independent migration runs.

The pod recreation baseline incurs the highest service interruption because recreating the UE pod requires complete reinitialization of the communication stack. Consequently, the switching delay exceeds $2$$~s$ and is accompanied by the largest user-plane recovery latency, packet loss, and jitter. Preserving the existing pod and restarting only the UE process substantially reduces the interruption, but service continuity is still affected by transient forwarding disruption, resulting in measurable packet loss and increased latency.

Orchra further reduces the interruption by transferring execution state rather than recreating network functions. The migration consists of state export, state import, PFCP reconfiguration, and N2 completion, which together complete in approximately $48$$~ms$. As a result, user-plane forwarding resumes almost immediately after the control-plane migration, requiring only about $2$$~ms$ before the first packet is observed on the destination UPF. Compared with the two baselines, Orchra therefore reduces the switching delay by more than an order of magnitude while avoiding costly UE reattachment and PDU session re-establishment procedures.

The reduction in interruption directly translates into improved service quality. Packet loss decreases from $9.45\%$ under pod recreation and $3.12\%$ under process restart to below $1\%$ with Orchra, while the transit jitter is reduced to approximately $1.2$$~ms$. These results indicate that preserving protocol state during migration enables rapid traffic resumption with minimal degradation of data-plane performance.

Finally, the state micro-benchmarks demonstrate that protecting migrated state with AES-GCM introduces only a modest computational overhead. The combined encryption and decryption penalty is approximately $1.3$$~ms$, which is negligible compared with the overall migration latency and therefore does not materially affect the end-to-end switching performance.

\begin{table*}[t]
    \centering
    \caption{\footnotesize Evaluation metrics for Orchra-assisted slice migration. Performance metrics are reported as mean $\pm$ standard deviation together with the 95\% confidence interval over $N=200$ independent experimental runs.}
    \label{table:evaluation_metrics}
    \rowcolors{3}{LightGray}{white}
    %\footnotesize
    \begin{tabularx}{\textwidth}{@{}lYYY@{}}
    \toprule
    \textbf{Metric Component} & \textbf{Baseline: Pod Recreate} & \textbf{Baseline: No Pod Recreate} & \textbf{Orchra: State Migration} \\
    \midrule
        \textbf{Control Plane Performance} & & & \\
        Export time ($t_{\mathrm{export}}$) [ms]       & -- & -- & $15.24 \pm 2.11$ [0.41] \\
        Import time ($t_{\mathrm{import}}$) [ms]       & -- & -- & $22.10 \pm 3.45$ [0.68] \\
        PFCP reconfiguration ($t_{\mathrm{pfcp}}$) [ms] & -- & -- & $8.15 \pm 1.02$ [0.20] \\
        N2 update latency ($t_{\mathrm{n2}}$) [ms]          & -- & -- & $2.86 \pm 1.84$ [1.73] \\
    \midrule
        \textbf{User Plane Performance} & & & \\
        User-plane recovery latency (First packet delay) ($t_{\mathrm{UP}}$) [ms]    & $245.50 \pm 32.10$ [6.29] & $112.40 \pm 15.80$ [3.10] & $1.86 \pm 1.57$ [1.73] \\
        Total switching delay ($t_{\mathrm{switch}}$) [ms]                    & $2099.17 \pm 33.29$ [6.52] & $181.26 \pm 18.11$ [3.55] & $48.35 \pm 10.67$ [3.46] \\
    \midrule
        \textbf{Data Plane Quality (SLA)} & & & \\
        Packet Loss Rate [\%]                         & $9.45 \pm 1.82$ [0.35]   & $3.12 \pm 0.65$ [0.12]    & $\mathbf{0.78 \pm 0.08}$ \textbf{[0.05]} \\
        Transit Jitter Delta ($\Delta J$) [ms]        & $15.40 \pm 4.12$ [0.81]   & $8.57 \pm 2.04$ [0.40]    & $\mathbf{1.22 \pm 0.18}$ \textbf{[0.03]} \\
    \midrule
        \textbf{State Micro-benchmarks (Security)} & & & \\
        Write Path (Plaintext) [ms]                   & -- & -- & $0.943 \pm 0.044$ [0.01] \\
        Write Path (AES-GCM Encrypted) [ms]           & -- & -- & $1.795 \pm 0.087$ [0.02] \\
        Read Path (Plaintext) [ms]                    & -- & -- & $0.621 \pm 0.035$ [0.01] \\
        Read Path (AES-GCM Decrypted) [ms]            & -- & -- & $1.036 \pm 0.061$ [0.01] \\
        \rowcolor{LightPink} Net Cryptographic Penalty [ms] & -- & -- & $\mathbf{+1.267 \pm 0.11}$ \textbf{[0.02]} \\
    \bottomrule
    \end{tabularx}
\end{table*}

\subsubsection{Security Overhead Analysis}
\begin{figure}[htbp]
    \centering
    \includegraphics[width=0.5\textwidth]{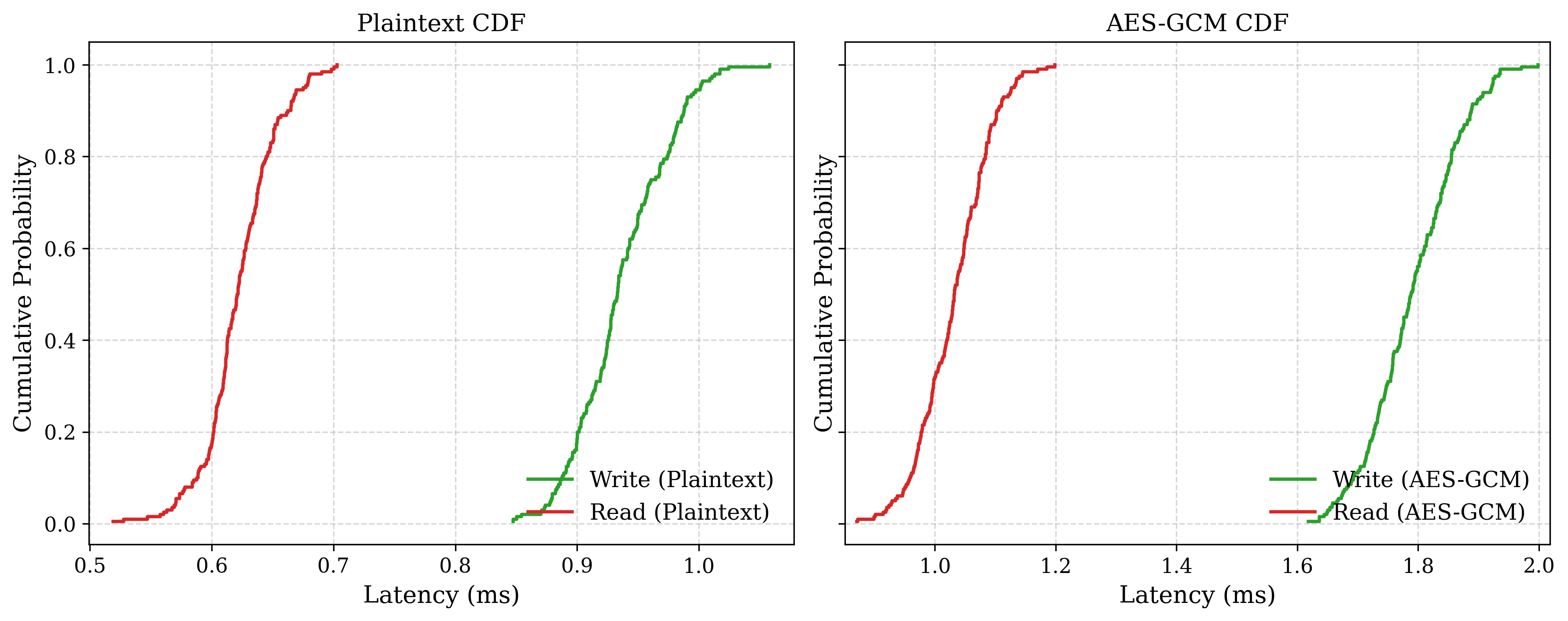}
    \caption{\footnotesize Empirical Cumulative Distribution Function (CDF) comparing control-plane reconfiguration latencies between unsecured plaintext context and encrypted context migrations.}
    \label{fig:crypto_overhead_cdf}
\end{figure}

\paragraph{Read/Write Throughput with AES-GCM Cryptography}
To isolate the performance penalty of our protection layering, Figure~\ref{fig:crypto_overhead_cdf} illustrates the cumulative probability of the control-plane latency ($t_{CP}$) with and without symmetric state encryption enabled. 

The empirical results in this comparison demonstrate that symmetric state encryption introduces a deterministic, highly bounded horizontal translation to the right of the baseline migration curve, confirming a constant additive overhead that does not introduce stochastic jitter or processing variance. At the 50th percentile ($p_{50}$), the complete end-to-end control-plane migration latency ($t_{CP}$) rises modestly from 13.5 ms to 14.8 ms, which aligns directly with the overall $+1.267 \pm 0.11$ ms net cryptographic penalty measured in the state micro-benchmarks. This performance delta remains stable across the upper bounds of the distribution ($p_{90}$), validating that the security framework scales reliably under tail-latency conditions. However, the micro-benchmarks reveal a notable asymmetric throughput trade-off: while the read/decryption path is highly efficient (increasing by only 0.415 ms), the active write/encryption path increases by $90.3\%$ (from $0.943 \pm 0.044$ ms to $1.795 \pm 0.087$ ms), identifying state serialization as a critical write-throughput bottleneck that requires parallelization to avoid CPU starvation on the SMF control plane during high-concurrency event bursts.

\begin{figure}[htbp]
    \centering
    \includegraphics[width=0.5\textwidth]{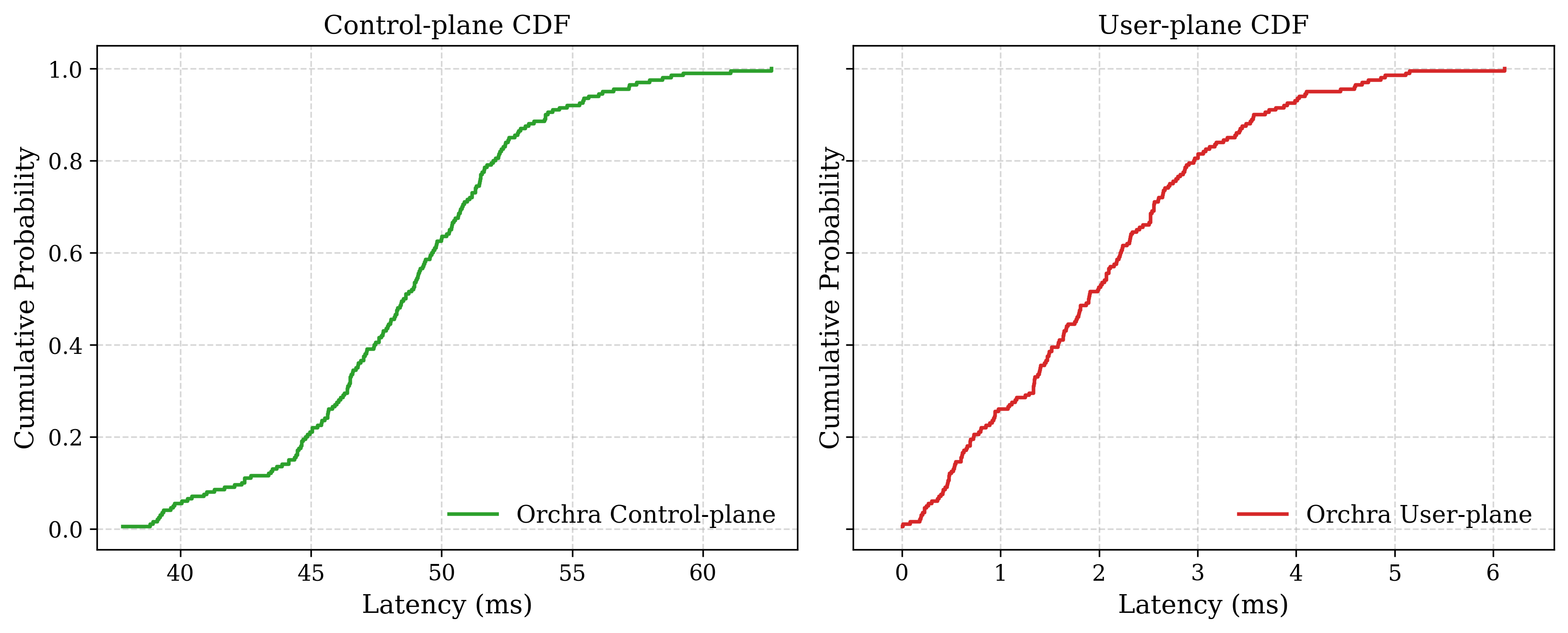}
    \caption{\footnotesize  Empirical CDF showing control-plane reconfiguration latency and user-plane transition delay.}
    \label{fig:latency_cdf}
\end{figure}

\paragraph{Control Plane vs. User Plane Migration Latency}
Figure~\ref{fig:latency_cdf} is a CDF that highlights the operational distinction between the Control Plane (signaling, orchestration, and state synchronization) and the User Plane (active packet disruption during a switchover). The User Plane curve rises sharply and vertically, demonstrating highly deterministic, low-jitter behavior where routing rule updates are applied within a tightly bounded, sub-millisecond range. In contrast, the Control Plane curve is horizontally translated to the right and features a more gradual slope, exposing the variable latency (jitter) inherent in non-deterministic cloud transactions, database interactions, and encryption overheads. Crucially, because the actual data-plane disruption window is kept exceptionally brief and decoupled from the longer control-plane orchestration path, the results confirm a successful implementation of Control-User Plane Separation (CUPS).

\section{Conclusion and Future Work}\label{sec:conclusions}
This paper presented Orchra, a stateful network slice migration framework that enables seamless migration of user equipment (UE) sessions across network slices without requiring repeated registration or PDU session re-establishment. By externalizing session state, coordinating control-plane migration, and re-anchoring the user plane, Orchra preserves session continuity while significantly reducing service interruption compared with conventional pod recreation and process restart approaches. Experimental evaluation on an OpenAirInterface-based 5G core demonstrates that the proposed framework reduces end-to-end switching latency to approximately $48~ms$, resumes user-plane forwarding within about 2 ms after control-plane completion, and maintains low packet loss and jitter. The results further show that protecting migrated state with authenticated encryption introduces only a modest processing overhead, making secure state migration practical for real-time operation.

Although these results demonstrate the feasibility of stateful slice migration, several challenges remain before such mechanisms can be widely deployed. Existing 3GPP and \ac{O-RAN} specifications do not yet define standardized procedures for migrating UE context across slices or administrative domains, limiting interoperability between implementations~\cite{oran2023}. In addition, maintaining consistent state across distributed control-plane functions during migration requires stronger coordination mechanisms to prevent partial updates or inconsistent network state in the presence of failures. As deployments scale to larger numbers of users and geographically distributed edge infrastructures, migration decisions will also need to become more adaptive in order to avoid unnecessary signaling and orchestration overhead~\cite{lee2024}. Finally, future work will extend the evaluation to large-scale, multi-tenant deployments by incorporating user mobility, heterogeneous transport networks, and fault scenarios to evaluate the robustness, scalability, and resilience of Orchra under realistic 5G edge operating conditions~\cite{Mukhtiar_etal, Chien_etal}. % 3gpp28541,~\cite{smith2023}

\section*{Acknowledgment}
\small
This work has been supported by the German Federal Ministry of Research, Technology and Space (BMFTR) within the projects \textit{SUSTAINET\_guarDian}\{16KIS2239K\} and \textit{Open6GHub+}\{16KIS2402K\}.

\begin{acronym}[HRTEM]
  \acro{QoS}{Quality of Service}
  \acro{QoE}{Quality of Experience}
  \acro{SDN}{Software Defined Networks}
  \acro{MEC}{Multi-access Edge Computing}
  \acro{UE}{User Equipment}
  \acro{5G}{Fifth Generation}UUID
  \acro{6G}{Sixth Generation}
  \acro{near-RT RIC}{Near-Runtime RAN Intelligent Controller}
  \acro{RIC}{RAN Intelligent Controller}
  \acro{PDU}{Protocol Data Unit}
  \acro{eMBB}{Enhanced Mobile Broadband}
  \acro{URLLC}{Ultra-Reliable and Low-Latency Communications}
  \acro{O-RAN}{Open Radio Access Network}
  \acro{OAI}{Open Air Interface}
  \acro{NWDAF}{Network Data Analytics Function}
  \acro{3GPP}{ Third Generation Partnership Project}
  \acro{AMF} {Access and Mobility Management Function}
  \acro{UDP} {User Datagram Protocol}
  \acro{CUPS}{ Control and User Plane Separation}
  \acro{DNS} {Domain Name System}
  \acro{FAR} {Forwarding Action Rule}
  \acro{FARs}{ Forwarding Action Rules}
  \acro{gNB} {g Node B (Next Generation Node B)}
  \acro{GTP-U} {GPRS Tunneling Protocol – User plane}
  \acro{HTTP} {Hypertext Transfer Protocol}
  \acro{NAS}{Non-Access Stratum}
  \acro{NF}{Network Function}
  \acro{NRF} {Network Repository Function}
  \acro{NSSAI} {Network Slice Selection Assistance Information}
  \acro{S-NSSAI}{Single Network Slice Selection Assistance Information}
  \acro{OAI}{OpenAirInterface}
  \acro{OAI-SMF} {OpenAirInterface Session Management Function}
  \acro{PDR}{Packet Detection Rule}
  \acro{PDRs}{Packet Detection Rules}
  \acro{PFCP}{Packet Forwarding Control Protocol}
  \acro{RAN}{Radio Access Network}
  \acro{REST}{Representational State Transfer}
  \acro{NSSAI}{Single Network Slice Selection Assistance Information}
  \acro{SBA}{Service-Based Architecture}
  \acro{SBI}{Service-Based Interface}
  \acro{SLA}{Service Level Agreement}
  \acro{SMF}{Session Management Function}
  \acro{TEID}{Tunnel Endpoint Identifier}
  \acro{UDM}{Unified Data Management}
  \acro{UDR}{Unified Data Repository}
  \acro{UE}{User Equipment}
  \acro{UPF}{User Plane Function}
  \acro{SEPP}{Security Edge Protection Proxy}
  \acro{AES-GCM}{Advanced Encryption Standard - Galois/Counter Mode}
  \acro{PLMN}{Public Land Mobile Network}
  \acro{H-PLMN}{Home-Public Land Mobile Network}
  \acro{V-PLMN}{Visitor-Public Land Mobile Network}
  \acro{mMTC}{Massive Machine-Type Communication}
  \acro{UUID}{Universally Unique Identifier}
  \acro{HTTP}{HyperText Transfer Protocol}
  \acro{AES}{Advanced Encryption Standard}
  \acro{CBC}{Cipher Block Chaining}
  \acro{HMAC}{Hash-based Message Authentication Code}
  \acro{SHA256}{Secure Hash Algorithm 256}
\end{acronym}
\bibliography{model.bib}{}

@techreport{3gpp_slice_orchestra,
  author = {3GPP},
  title = {3GPP TS 28.531 V16.14.1 (2023-01): Management and Orchestration; Provisioning},
  institution = {3rd Generation Partnership Project (3GPP)},
  type = {Technical Specification},
  number = {TS 28.531},
  year = {2023},
  month = {09},
  note = {Version 16.14.1, Release 16},
  url = {https://www.3gpp.org/dynareport/28531.htm}
}

@article{sajjad2020interslice,
  author       = {Muhammad Mohtasim Sajjad and Carlos J. Bernardos and Dhammika Jayalath and Yu{-}Chu Tian},
  title        = {{Inter‐Slice Mobility Management in 5G: Motivations, Standard Principles, Challenges and Research Directions}},
  journal      = {{IEEE} Communications Standards Magazine},
  volume       = {6},
  number       = {2},
  pages        = {93--100},
  year         = {2022},
  doi          = {10.1109/MCOMSTD.0001.2000025},
  url          = {https://doi.org/10.1109/MCOMSTD.0001.2000025},
}

@techreport{oran2023,
  author      = {{European Telecommunications Standards Institute}},
  title       = {Publicly Available Specification (PAS); O-RAN Slicing Architecture (O-RAN.WG1.Slicing-Architecture-R003-v11.00)},
  institution = {ETSI},
  type        = {Technical Specification},
  number      = {ETSI TS 104 041 V11.0.0},
  month       = {03},
  year        = {2025},
  url         = {https://www.etsi.org/deliver/etsi_ts/104000_104099/104041/11.00.00_60/ts_104041v110000p.pdf}
}

@article{lee2024,
  author       = {Lee, J. and Zhang, X. and Huang, Y. and others},
  title        = {OnSlicing: Online End-to-End Network Slicing with Reinforcement Learning},
  journal      = {IEEE/ACM Transactions on Networking},
  volume       = {32},
  number       = {4},
  year         = {2024},
  publisher    = {IEEE/ACM}
}

@inproceedings{Mukhtiar_etal,
  author = {Ahmad, Mukhtiar and Bilal, Faaiq and Ali, Mutahar and Nawazish, Syed Muhammad Ali and Salman, Amir and Ali, Shazer and Ahmad, Fawad and  Qazi, Zafar Ayyub},
  title = {Warping the Edge: Enabling Instant Mobility for Stateful Applications over 5G and Beyond},
  year = {2025},
  isbn = {9798400722387},
  publisher = {Association for Computing Machinery},
  url = {https://doi.org/10.1145/3769102.3770610},
  doi = {10.1145/3769102.3770610},
  booktitle = {Proceedings of the Tenth ACM/IEEE Symposium on Edge Computing},
  articleno = {25},
  numpages = {18},
  series = {SEC '25}
}

@article{Chien_etal,
  title = {End-to-end slicing as a service with computing and communication resource allocation for multi-tenant 5G systems},
  author = {Chien, \{Hsu Tung\} and Ying-Dar Lin and Lai, \{Chia Lin\} and Wang, \{Chien Ting\}},
  note = {Publisher Copyright: {\textcopyright} 2019 IEEE.},
  year = {2019},
  doi = {10.1109/MWC.2019.1800466},
  volume = {26},
  pages = {104--112},
  journal = {IEEE Wireless Communications},
  issn = {1536-1284},
  publisher = {Institute of Electrical and Electronics Engineers Inc.},
  number = {5}

}

@article{mozaic5g,
    author = {Nikaein, Navid and Chang, Chia-Yu and Alexandris, Konstantinos},
    title = {Mosaic5G: agile and flexible service platforms for 5G research},
    year = {2018},
    issue_date = {July 2018},
    publisher = {Association for Computing Machinery},
    volume = {48},
    number = {3},
    issn = {0146-4833},
    url = {https://doi.org/10.1145/3276799.3276803},
    doi = {10.1145/3276799.3276803},
    journal = {SIGCOMM Comput. Commun. Rev.},
    month = {09},
    pages = {29–34},
    numpages = {6}
}

@article{kaltenberger2020openairinterface,
  title={OpenAirInterface: {A}n open-source system for simulation, emulation, and real-time experimentation of {5G} cellular networks},
  author={Kaltenberger, Florian and others},
  journal={IEEE Communications Magazine},
  volume={58},
  number={7},
  pages={54--60},
  year={2020}
}

@INPROCEEDINGS{Gallego,
  author={Vázquez-Gallego, Francisco and Nasreddine, Jad and Carmona-Cejudo, Estela and Murillo, Yuri and Vilalta, Ricard and Veyssiere, Philippe and Dalgitsis, Michail and Serrano, María A. and Antonopoulos, Angelos and Polo, Javier and Bastida, Judit and Luque, José López},
  booktitle={2023 IEEE Future Networks World Forum (FNWF)}, 
  title={Cross-Border 5G Seamless Connectivity for Connected and Automated Mobility: Challenges, Network Implementation, and Lessons Learnt}, 
  year={2023},
  volume={},
  number={},
  pages={1-6},
  doi={10.1109/FNWF58287.2023.10520360}
}

@article{kousaridas20215g,
  title={5g vehicle-to-everything services in cross-border environments: Standardization and challenges},
  author={Kousaridas, Apostolos and Fallgren, Mikael and Fischer, Edwin and Moscatelli, Francesca and Vilalta, Ricard and M{\"u}hleisen, Maciej and Barmpounakis, Sokratis and Vilajosana, Xavier and Euler, Sebastian and Tossou, Bruno and others},
  journal={IEEE Communications Standards Magazine},
  volume={5},
  number={1},
  pages={22--30},
  year={2021},
  publisher={IEEE}
}

@INPROCEEDINGS{Marquez_etal,
  author={Marquez-Barja, Johann and Hadiwardoyo, Seilendria A. and Maglogiannis, Vasilis and Naudts, Dries and Moerman, Ingrid and Hellinckx, Peter and Verbrugge, Sofie and Delaere, Simon and Vandenberghe, Wim and Kenis, Eric and Campodonico, Maria Chiara and Kusumakar, Rakshith and Meines, Job and Vandenbossche, Joost},
  booktitle={2021 IEEE 18th Annual Consumer Communications and Networking Conference (CCNC)}, 
  title={Enabling cross-border tele-operated transport in the 5G Era: The 5G Blueprint approach}, 
  year={2021},
  volume={},
  number={},
  pages={1-4},
  doi={10.1109/CCNC49032.2021.9369619}
}

@ARTICLE{afolabi2018network,
  author={Afolabi, Ibrahim and Taleb, Tarik and Samdanis, Konstantinos and Ksentini, Adlen and Flinck, Hannu},
  journal={IEEE Communications Surveys and Tutorials}, 
  title={Network Slicing and Softwarization: A Survey on Principles, Enabling Technologies, and Solutions}, 
  year={2018},
  volume={20},
  number={3},
  pages={2429-2453},
  doi={10.1109/COMST.2018.2815638}
}

@inproceedings{saad2021slice,
  title={Inter-slice mobility management in 5G: motivations, standard principles, challenges, and research directions},
  author={Sajjad, Muhammad Mohtasim and Bernardos, Carlos J and Jayalath, Dhammika and Tian, Yu-Chu},
  journal={IEEE Communications Standards Magazine},
  booktitle ={IEEE Communications Standards Magazine},
  volume={6},
  number={1},
  pages={93--100},
  year={2022},
  publisher={IEEE}
}

@techreport{3gpp.29.500,
  author      = {{3GPP}},
  title       = {5G System; Technical Realization of Service Based Architecture; Stage 3},
  type        = {Technical Specification (TS)},
  number      = {29.500},
  institution = {3rd Generation Partnership Project},
  year        = {2023},
  month       = {09},
  note        = {version 16.15.0 Release 16 },
  url         = {https://www.etsi.org/deliver/etsi_ts/129500_129599/129500/16.15.00_60/ts_129500v161500p.pdf}
}

@article{taleb2019cross,
  author={Taleb, Tarik and Afolabi, Ibrahim and Samdanis, Konstantinos and Yousaf, Faqir Zarrar},
  journal={IEEE Network}, 
  title={On Multi-Domain Network Slicing Orchestration Architecture and Federated Resource Control}, 
  year={2019},
  volume={33},
  number={5},
  pages={242-252},
  doi={10.1109/MNET.2018.1800267}
}

@techreport{p1sec2026sepp,
  title       = {5G; Security assurance specification (SCAS) for the 5G Core Network (5GC) Network Product Class},
  author      = {ETSI},
  institution = {European Telecommunications Standards Institute (ETSI)},
  type        = {Technical Specification},
  number      = {ETSI TS 133 517 V16.1.0},
  year        = {2020},
  month       = {10},
  url         = {https://www.etsi.org/deliver/etsi_ts/133500_133599/133517/16.01.00_60/ts_133517v160100p.pdf}
}

@inproceedings{Nguyen_etal,
  author = {Nguyen, Huy-Trung and Nguyen, Ngoc-Quan and Le, Viet H and Tran, Sang D and Hoang, Linh D},
  title = {PROPOSED DATA ROAMING NETWORK SIMULATION ENVIRONMENT FOR 5G NETWORK},
  year = {2026},
  isbn = {9798400721250},
  publisher = {Association for Computing Machinery},
  url = {https://doi.org/10.1145/3785520.3785539},
  doi = {10.1145/3785520.3785539},
  booktitle = {Proceedings of the 2025 10th International Conference on Cloud Computing and Internet of Things},
  pages = {138–147},
  series = {CCIOT '25}
}

@INPROCEEDINGS{Sajjad,
  author={Sajjad, Muhammad Mohtasim and Jayalath, Dhammika and Tian, Yu-Chu and Bernardos, Carlos J.},
  booktitle={2020 IEEE 31st Annual International Symposium on Personal, Indoor and Mobile Radio Communications}, 
  title={On Session Continuation among Slices for Inter-Slice Mobility Support in 3GPP Service-based Architecture}, 
  year={2020},
  volume={},
  number={},
  pages={1-7},
  doi={10.1109/PIMRC48278.2020.9217332}
}

@ARTICLE{Mohammedali,
  author={Mohammedali, Noor A. and Kanakis, Triantafyllos and Agyeman, Michael Opoku and Al-Sherbaz, Ali},
  journal={IEEE Access}, 
  title={A Survey of Mobility Management as a Service in Real-Time Inter/Intra Slice Control}, 
  year={2021},
  volume={9},
  number={},
  pages={62533-62552},
  doi={10.1109/ACCESS.2021.3074024}
}

@inproceedings{sakic2020decoupling,
  title={Decoupling of distributed consensus, failure detection and agreement in SDN control plane},
  author={Sakic, Ermin and Kellerer, Wolfgang},
  booktitle={2020 IFIP Networking Conference (Networking)},
  pages={467--475},
  year={2020},
  organization={IEEE}
}

@techreport{3gpp_38_300,
  author      = {{3GPP}},
  title       = {{NR; NR and NG-RAN Overall description; Stage-2}},
  type        = {Technical Specification (TS)},
  institution = {{3rd Generation Partnership Project (3GPP)}},
  number      = {38.300},
  url         = {https://www.3gpp.org/dynareport/38300.htm},
  note        = {Version 16.0.0, Release 16},
  year        = {2020}
}
\bibliographystyle{IEEEtran}

\end{document}